\pdfoutput=1

\documentclass{amia}
\usepackage[labelfont=bf]{caption}
\usepackage{color}

\usepackage[super,comma]{natbib}
\usepackage{times}
\usepackage{soul}
\usepackage{url}
\usepackage[hidelinks]{hyperref}
\usepackage[utf8]{inputenc}
\usepackage{amsmath}
\usepackage{amsthm}
\usepackage{booktabs}
\usepackage{algorithm}
\usepackage{algorithmic}
\usepackage{multirow}
\usepackage{subcaption}
\usepackage{makecell}
\usepackage{amsfonts}
\usepackage{amssymb}
\usepackage{graphicx}
\usepackage{wrapfig}

\begin{document}

\setlength{\bibsep}{0.1pt}

\title{Cross-Vendor CT Image Data Harmonization Using CVH-CT}
\author{
Md Selim$^{1,3}$,
Jie Zhang, PhD$^2$,
Baowei Fei, PhD$^{5, 6}$,
Guo-Qiang Zhang, PhD$^7$ ,
Gary Yeeming Ge$^2$ ,
Jin Chen, PhD$^{1,3,4}$}

\institutes{
$^1$Department of Computer Science 
$^2$Department of Radiology 
$^3$Institute for Biomedical Informatics 
$^4$Department of Internal Medicine,  University of Kentucky, Lexington, KY 
$^5$Department of Bioengineering, University of Texas at Dallas, Richardson, TX 
$^6$Department of Radiology, UT Southwestern Medical Center, Dallas, TX
$^7$Department of Neurology, University of Texas Health Science Center at Houston, Houston, TX 
}

\maketitle

\noindent{\bf Abstract}
\textit{
While remarkable advances have been made in Computed Tomography (CT), most of the existing efforts focus on imaging enhancement while reducing radiation dose. How to harmonize CT image data captured using different scanners is vital in cross-center large-scale radiomics studies but remains the boundary to explore.
Furthermore, the lack of paired training image problem makes it computationally challenging to adopt existing deep learning models. 
We propose a novel deep learning approach called CVH-CT for harmonizing CT images captured using scanners from different vendors. The generator of CVH-CT uses a self-attention mechanism to learn the scanner-related information. We also propose a VGG feature based domain loss to effectively extract texture properties from unpaired image data to learn the scanner based texture distributions.
%
The experimental results show that CVH-CT is clearly better than the baselines because of the use of the proposed domain loss, and CVH-CT can effectively reduce the scanner-related variability in terms of radiomic features. 
}

\section{Introduction}
%
Computed Tomography (CT) is one of the most commonly used imaging modalities for patient diagnostics due to its ability to capture detailed anatomical features~\cite{prince2006medical}. For years, Canon (former Toshiba), Siemens, General Electric (GE), and Philips have been supplying high-quality CT scanners worldwide. Owing to increasing competition frequent innovations regarding slice count, dose optimization, reconstruction methods, etc. are taking place in the market. Each medical imaging vendor gradually develops its unique techniques~\cite{chirra2018empirical} to advance CT imaging. The divergence among CT imaging techniques, however, introduces image feature variations in terms of radiomic patterns~\cite{chirra2018empirical,ours_aamp}, which greatly hinder the progress of cross-vendor data sharing, ambiguity in large-scale data analysis, and automated diagnostics~\cite{Multi-Stakeholder-Participation-Buckler}. 

To address the radiomic feature discrepancy problem, one of the computational approaches is to normalize the radiomic features of CT images captured with different protocols. However, since the concepts of radiomic features are not well defined~\cite{foy2018variation}, an image feature normalization tool that works for a set of radiomic features may not work for the others. Alternatively, we can harmonize CT image data directly while preserving their anatomic details~\cite{ours_aamp,selim2020stan}. From the harmonized images, radiomics analysts can extract their desired radiomic features for further analysis without worrying about the feature discrepancy problem. Mathematically, let $x$ be a CT image acquired using a scanner, $\hat{x}$ be its corresponding image captured with a different scanner, the image harmonization framework aims to compose a synthetic image $x'$ from $x$, such that $x'$ follows the feature distributions of $\hat{x}$ rather than $x$.

%
The CT image harmonization problem can be viewed either within scanner or between scanners. Recent progress on this topic has been focused on the former one. 
Choe et al~\cite{choe2019deep} developed a Convolutional Neural Network (CNN)-based approach for CT image standardization. The model learns the residual representation of the target images, and then a residual image is combined with its source image to generate a synthesized image. The model, since it trains a CNN from scratch, requires large training data.
Liang et al~\cite{liang2018ganai} proposed a cGAN-based~\cite{pix2pix} CT image standardization model named GANai. An alternative training strategy was developed to effectively learn the data distribution. GANai achieved better performance comparing with cGAN and the traditional histogram matching approach~\cite{gonzalez2012digital}. However, GANai focuses on the relatively easier image patch synthesis problem rather than the whole DICOM image synthesis problem.
Selim et al~\cite{selim2020stan} proposed a GAN-based CT image standardization model named STAN-CT. In STAN-CT, a loss function was developed to consider both the latent space loss and the feature space loss. While the former is adopted for the generator to establish a one-to-one mapping from standard images to synthesized images, the latter allows the discriminator to critic the texture features of both standard and synthesized images. Similar to GANai, STAN-CT was applied at image patches and only a few texture features were used as the evaluation criteria.
Another GAN-based CT image standardization model named RadiomicGAN~\cite{selim2021radiomic} took the advantage of transfer learning and proposed a dynamic window-based training approach to adopt the learned information from the RGB image domain into the CT image domain. The results were evaluated on a wide range of radiomics features.

%
%
However, all these models require paired training data, greatly limiting the application scope to harmonizing images captured with the same type of scanners. 

\begin{figure}
\centering
\includegraphics[width=\columnwidth]{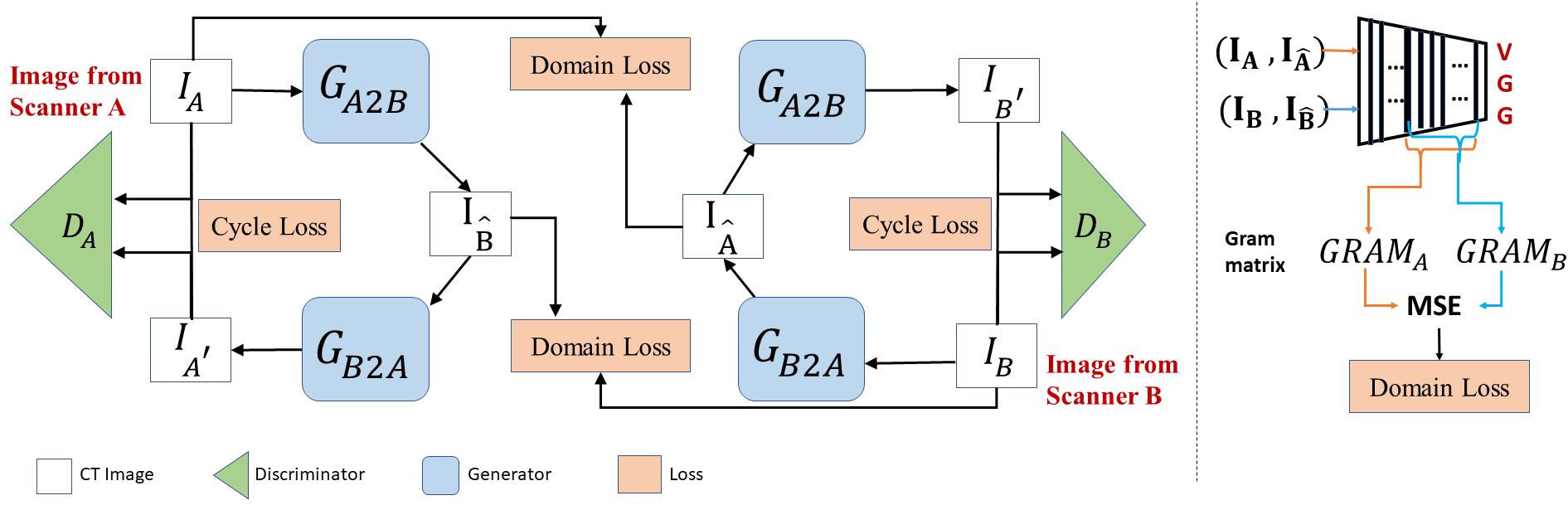}
\caption{\textbf{Architecture of CVH-CT.} The generators $A2B$ and $B2A$ are both CNN networks with a similar structure. The discriminator D is a fully convolutional network classifier for determining the source of the synthesized images.  A VGG texture feature-based domain loss is used to train the model with unpaired data.  Domain loss is calculated as a mean square error between the gram matrix representation of $I_A$ and $I_B$.} \label{fig:model}
\end{figure}

The cross-vendor CT image harmonization remains a critical bottleneck for inter-institutional data harmonization. This is mainly because it is difficult to obtain paired imagery data ~\cite{you2019ct}. For example, a patient is scanned using a GE scanner, it is less likely that the patient will be scanned using a Siemens scanner in close time. 
%
%
In this paper, we present a novel deep learning model, called CVH-CT (cross-vendor harmonization of CT images), for cross-vendor CT image harmonization  (see Figure~\ref{fig:model}). CVH-CT relaxes the need for paired training data. Using unpaired training data, CVH-CT can synthesize images from vendor A to vendor B and vise versa. CVH-CT integrates a self-attention layer named CBAM (Convolutional Block Attention Map)~\cite{woo2018cbam} to systematically learn the global features that appear in the images due to the use of different vendors. CVH-CT borrows the CycleGAN~\cite{CycleGAN2017} loss and improves it with the feature-based domain loss to determine the feature gap between the synthesized images and the target images in the target domain. Overall, CVH-CT has the following advantages:

\begin{enumerate}
	\item CVH-CT effectively learns the feature distributions between two different CT image domains without paired training data.
	\item The self-attention block, CBAM, with the convolutional layers helps the network to capture the scanner's relevant global distribution.
	\item The domain loss can calculate the feature gap between unpaired data which assists the network to learn the target domain distribution.
	\item Experimental results show CVH-CT is better than the state-of-the-art CT image harmonization methods.
\end{enumerate}

\section{Background}
\noindent{\bf CT Scanner. }
CT is a widely used clinical imaging modality for patient diagnostics. CT scanner vendors in the market, including Canon, Siemens, GE, and Philips, have developed unique reconstruction algorithms to improve image quality while reducing the dosage~\cite{chirra2018empirical}. The CT image radiomic features discrepancy problem due to the use of scanners with different imaging techniques poses a gap between CT imaging and radiomics studies~\cite{chirra2018empirical,ours_aamp}. 


\noindent{\bf Radiomic Feature. }
Image features, commonly known as the radiomic features, are critical for radiomics study, e.g. tumor characterization~\cite{yip2016applications}. Mathematical and statistical models are used to extract these features from images. 
Radiomic features reflect the cellular and genetic levels phenotypic patterns that are hidden from the naked eye~\cite{ yang2011quantifying,yip2016applications}. Thus, there is a great potential to capture tumor heterogeneity and phenotypic details with radiomic features. 
However, the effectiveness of radiomic features, especially for large-scale cross-institute studies, is greatly reduced due to the non-standard practice of medical image acquisition~\cite{berenguer2018radiomics}, since radiomic features are dependent on both inter and intra-scanner protocol settings~\cite{hunter2013high, berenguer2018radiomics}. 

\noindent{\bf Image synthesis. }
Image synthesis is a common practice for artificial data generation which mimics the real data distribution~\cite{huang2018introduction}. An image synthesis algorithm may require no input or a random noise for data synthesis but this type of algorithm has less control on the synthesized data~\cite{hall1983testbed}. Medical data synthesis requires high quality with valid clinical meaning. So not all the available image synthesization models may not be appropriate for the medical domain. Recent progress on deep learning for image-to-image transformation provides better control on the synthesized data which makes a suitable choice to adopt it in the medical domain~\cite{yu2019retinal}. 

Generative Adversarial Networks (GAN), which are often used for data and image synthesis~\cite{huang2018introduction}, normally consist of a generator $G$ and a discriminator $D$. The generator that could be a Convolutional Neural Network (CNN) is responsible for generating fake data from noise, and the discriminator tries to identify whether its input is drawn from the real or fake data. Among all the GAN models, cGAN is capable of synthesizing new images based on a prior distribution~\cite{cgan}. 
The conventional GAN-based image-to-image translation model training requires paired training data that limits its scope where paired data is unavailable. A variation of GAN named CycleGAN~\cite{CycleGAN2017} is designed by combining two GAN models to synthesized images for unpaired data.

\section{Methods}
CVH-CT is designed to harmonize the scanner-related variability by generating synthesized CT images without the need for paired cross-vendor training data.
%
%
Built on top of the CycleGAN framework, CVH-CT incorporates consecutive convolutional and self-attention layers to learn the global features of entire CT images efficiently. CVH-CT also introduces a new VGG~\cite{simonyan2014very} feature-based domain loss function, which can be directly calculated using unpaired images. 

\subsection*{CVH-CT Architecture}\label{sec:GAN}
Based on the cycle generative adversarial network method called CycleGAN, CVH-CT consists of two generative adversarial networks (GANs). Each GAN has its generator $G$ and discriminator $D$ but with the same structure. A GAN tries to map the data distribution from data domain $A$ to domain $B$. The two generators are $G_{A2B}: A \rightarrow B$ and $G_{B2A}: B \rightarrow A$. The discriminator $D_A$ is responsible to distinguish between the data being real or fake in terms of domain A and the same job is done by $D_B$ for domain B. 
In Figure~\ref{fig:model}, given training datasets from two domains, e.g. one set of CT images captured using scanner A and another set captured using scanner B, an initial mapping of $G_{A2B}$ is learned to be able to generate synthesized of domain B based on the input image from A. The mapping of $G_{B2A}$ is learned to generate synthesized domain A's image using the input image from B. Discriminator $D_A$ tries to determine whether the input image is from domain $A$ or not and the same for discriminator $D_B$. 

\begin{figure}
\includegraphics[width=\textwidth]{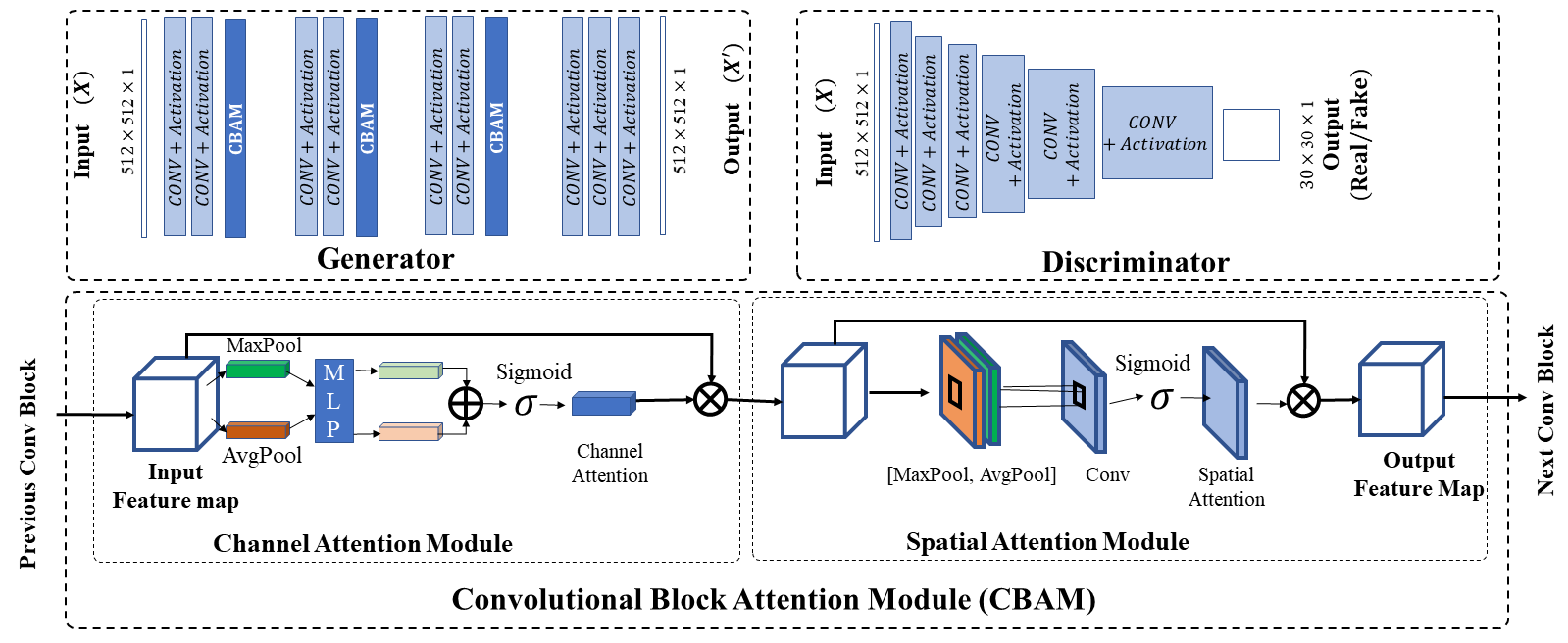}
\caption{{\bf The architecture of CVH-CT generator and discriminator.} The generator use a series of convolutional layers followed by a Convolutional Block Attention Map (CBAM). The convolutional layers are responsible to fetch the local details while the CBAM is responsible to fetch the global details within the feature-map coming from its previous convolutional layers. } \label{fig:framework}
\end{figure}

\noindent {\bf CVH-CT Generator. }
The generator of CVH-CT, as shown in Figure~\ref{fig:framework}, is a convolutional neural network with dedicated convolutional-attention layers. Specifically, each convolutional layer followed by an activation function is used to extract features from the previous layer and store it in a feature map. The convolutional layer dimension is kept consistent throughout the network. For every two convolutional layers (except the last three layers), a self-attention mapping is constructed to capture key global features from the corresponding convolutional layers. These self-attention layers are designed to learn the global features from the spatial dimension and the inter-channel dimension. 

\noindent {\bf Convolutional Block Attention Map (CBAM). } 
In complementary to the convolutional filters focused on extracting local features, self-attention layers are added into CVH-CT to capture the common features in each channel of the network. Such network structure change allows CVH-CT to capture domain-relevant features in the cross-vendor CT image harmonization problem. Figure~\ref{fig:framework} illustrates the architecture of CVH-CT where each self-attention layer called Convolutional Block Attention Map (CBAM)~\cite{woo2018cbam} consists of the channel-wise attention and the spatial attention. 




\textit{Channel-wise Attention Map: }
The channel-wise attention map is the first phase of CBAM that establishes an inter-channel relationship within a feature map. Max-pooling and avg-pooling operations are used to extract features from the spatial dimensions. Both feature maps are passed through a Multilayer Perceptron neural network (MLP). The output features are concatenated, and an element-wise product operation is used to produce the channel attention map.

\textit{Spatial Attention Map: }
The spatial attention map works on top of the output from the channel-wise attention map. Avg-pooling and max-pooling are used along with the channel axis. The average and maximum feature values pool out the impotent feature. Convolution is done on the concatenated feature of the average and max-pooing feature. After softmax, the output is element-wise multiplied with the input feature map. The output feature map is passed to the next layer for further process.

\subsection*{CVH-CT Loss}\label{sec:GAN}
Conventional CycleGAN is trained using a comprehensive loss function including adversarial loss, cycle consistency loss, and identity loss~\cite{CycleGAN2017}. In CVH-CT, we introduce a new loss function called ``domain loss'' to compare key texture features between unpaired synthesized and target domain images.

\noindent {\bf Domain Loss. }
It is a widely used practice to extract style and content features using a pre-trained VGG network~\cite{gatys2016image}. In the cross-vendor CT image harmonization task, a synthesized image and an unpaired target image may have different anatomic structures but similar scanner-relevant texture features (aka. style). We adopted the style-related features from the trained VGG network, which are traditionally the layers closer to the fully connected layers in VGG. The domain loss is calculated as a Mean Square Error (MSE) of the gram-matrix representation of those layers~\cite{gatys2016image}. The domain loss is defined as:
\begin{equation}
\mathcal{L}_{f_A} = \frac{1}{m}\sum_{i=1}^{m} \left(  GRAM(G_{A2B}(a_i))-GRAM(b_i)  \right)^2
\end{equation}
\begin{equation}
\mathcal{L}_{f_B} = \frac{1}{m}\sum_{i=1}^{m} (  GRAM(G_{B2A}(b_i))-GRAM(a_i)  )^2
\end{equation}
\begin{equation}
\mathcal{L}_f = \mathcal{L}_{f_A}+\mathcal{L}_{f_B}
\end{equation}
where $GRAM(.)$ is the gram-matrix calculated by multiply the feature maps by itself as described in Gatys et al~\cite{gatys2016image}. A and B are two image domains, and m is the batch size.

\noindent {\bf Adversarial loss. }
We apply the adversarial loss to both generators, where each generator synthesizes images of its target domain, while its corresponding discriminator tries to distinguish its input images are from the target domain distribution or not. In  model training, the generator aims to minimize the adversarial loss while its corresponding discriminator tries to maximize it. Note the discriminators if CVH-CT are trained only with $\mathcal{L}_{adv(D)}$.

\begin{equation} \label{eq:wgan_d}
\mathcal{L}_{adv({G_{A2B}},D_B,A)} = \frac{1}{m}\sum_{i=1}^{m} (1-D_B({G_{A2B}}(a_i)))^2
\end{equation}
\begin{equation} \label{eq:wgan_d}
\mathcal{L}_{adv({G_{B2A}},D_A,B)} = \frac{1}{m}\sum_{i=1}^{m} (1-D_A({G_{B2A}}(b_i)))^2
\end{equation}
\noindent where $m$ is the batch size, $a\in A$ is the input image of domain $A$ and $b \in B$ is the input image from domain $B$.

\noindent {\bf Cycle Consistency Loss. }
The intuition of Cycle Consistency Loss is that if an input image $a_i$ goes through a generator and back again in its compliment generator by creating a close cycle then the final synthesized image will be the same as input image $a_i$.
Thus, the Cycle Consistency Loss is calculated as the L1 loss between the input image and the final synthesized image, and is calculated in two directions:
\begin{equation} \label{eq:f_loss_d}
  \mathcal{L}_{cyc} (G_{A2B},G_{B2A},A,B) = \frac{1}{m} \sum_{1}^{m} || G_{B2A}(G_{A2B}(a_i))-a_i|| + || G_{A2B}(G_{B2A}(b_i))-b_i||
\end{equation}

\noindent {\bf Identity Loss. } 
The identity loss is introduced to preserve the key details of the input and target domains. It is calculated by proving the target domain data as input of the generator and calculating the L1 loss between input and synthesized images. 
\begin{equation} \label{eq:f_loss_d}
  \mathcal{L}_{Idt} (G_{A2B},G_{B2A},A,B) = \frac{1}{m} \sum_{1}^{m} || G_{A2B}(b_i)-b_i|| + || G_{B2A}(a_i)-a_i||
\end{equation}

Finally, the total loss of the CVH-CT generator  $\mathbb{L}(G)$ is defined as:
\begin{equation}
\mathbb{L}(G) = \mathcal{L}_{adv(D)} + \lambda_1 \mathcal{L}_{cyc}+ \lambda_2 \mathcal{L}_{Idt} + \lambda_3 \mathcal{L}_{f}
\end{equation}
where $\lambda_1 \in [0,1]$, $\lambda_2 \in [0,1]$ and $\lambda_3 \in [0,1]$ are wight factors.

\section{Experimental Results} \label{sec_experiment}
\subsection*{Dataset} 
A multipurpose lungman chest phantom was scanned using the Siemens CT Somatom Force scanner and General Electric (GE) Revolution EVO scanner machine to obtain the training and testing data. The phantom has synthetic nodules in its lung which makes it an accurate life-size anatomical model of a human torso. Note that the phantom does not contain lung parenchyma, so the space within the vascular structure is filled with air. 
Using the human chest phantom with several synthetic lung nodules, in total 11,070 image slices were obtained from a GE scanner. In total 9,156 image slices were obtained using a Siemens scanner where 7,290 image slices are from the chest phantom and 1,866 are from lung cancer patients. All the image slices are $512 \times 512$ with 16-bit pixel encoding. The image acquisition parameters and the data description are in Table~\ref{tab:data}. Images in the training dataset are randomly paired. 
%
%
To evaluate the performance of CVH-CT and baseline models, we prepared a slice-by-slice paired dataset using phantom scans.  Using the GE and Siemens scanners, in total 250 image pairs were generated with five different KVPs and a fixed (5 mm) slice thickness.

Also, we tested the performance of CVH-CT on CT images collected from the same scanner but with two different image reconstruction kernels using patient scans.
%
The model was trained with a total of 9,580 CT image slices from lung cancer patients using four different slice thicknesses (0.5, 1, 1.5, 3mm) using the Siemens CT Somatom Force scanner. To enable model performance evaluation, we reconstructed two images from the same scan using reconstruction kernels Bl64 and Br40 respectively. Bl64 is the reconstruction kernel often used for lung screening, and Br40 is used for regular screening.

\begin{table}
\footnotesize
\centering
\caption{Description of CT images acquired from GE and Siemens scanners. }
	\begin{tabular}{| c | c | c | }
	\hline

\textbf{CT scanner}	&	GE Revolution EVO	&	Siemens CT Somatom Force		\\ \hline
\textbf{Reconstruction Kernel}	&	Lung	&	Bl64		\\ \hline
\textbf{Slice Thickness (mm)}	&	0.625, 1.25, 2.5, 3.75 and 5	&	1,1.5, 3 and 5		\\ \hline
\textbf{KVP}	&	70,80,100,120 and 140	&	70,80,100,120 and 140		\\ \hline
\textbf{Total No. of Slices}	&	11,070 (phantom)	&	7,290 (phantom), 1,866 (human)		\\ \hline

	\end{tabular} \label{tab:data}
\end{table}

\subsection*{Model Implementation} 
The generator of CVH-CT consists of seven convolutional layers and three CBAM blocks. Each convolutional layers have 64 filters with kernel size $4\times 4$ and stride=1.
The discriminator is a fully connected convolutional neural network with seven hidden layer~\cite{pix2pix}. The convolutional layers used $4 \times 4$ filters. LeakyRelu~\cite{xu2015empirical} is adopted as the activation function in all the hidden layers. The last layer of the generators uses $Tanh$ activation and discriminators use $Sigmoid$ activation. 
The model uses $80\times 80$ soft-tissue image patches from the training dataset to train the model and the testing is done using the full CT slice that is $512\times 512$. The synthesization is done with the Hounsfield Unit (HU) ranging from $-1000$ to $900$. 
Random weights are used during the network initialization phase. Maximum training epochs are set to 50 with the learning rate being 0.0001 with momentum 0.5. The batch size is set to 32. 

CVH-CT is implemented in PyTorch and runs on a Linux computer server with eight Nvidia GTX 1080 GPU cards. It takes about 4 hours to train the model from scratch. Once the model is trained, it takes about two seconds to harmonize a CT image slice. (Source code: https://github.com/AVAILABLE-SOON)

\subsection*{Evaluation Metric}
Model performance was evaluated systematically at the whole image level and with randomly selected regions of interest (ROIs) from the soft tissues. 
For each CT image or ROI, quantitative radiomic features are extracted from six different feature classes using Pyradiomics~\cite{van2017computational}. 
These feature classes and the number of features are, First Order Statistics (18), Gray Level Co-occurrence Matrix (GLCM, 24), Gray Level Run Length Matrix (GLRLM, 16), Gray Level Size Zone Matrix (GLSZM, 16), Neigbouring Gray Tone Difference Matrix (NGTDM, 5) and Gray Level Dependence Matrix (GLDM, 14). The six feature classes contain a total of 93 features.

We evaluated the CT image harmonization models by the radiomic features reproducibility analysis and the visual quality of the synthesized images. 
The radiomic feature reproducibility is analyzed by Concordance Correlation Coefficient (CCC)~\cite{lawrence1989concordance} score and the visual quality of the synthesized images are determined by Peak Signal-to-Noise Ratio (PSNR)~\cite{korhonen2012peak}, Structural SIMilarity (SSIM)~\cite{wang2004image}, and Normalized Cross Correlation (NCC)~\cite{zhao2006image}. 

Concordance Correlation Coefficient (CCC, see Eq.~\ref{eq:ccc}) was employed to measure the level of reproducibility of radiomic features~\cite{choe2019deep}. Mathematically, CCC represents the correlation between the input and the target image features in the six feature classes: 
\begin{equation}\label{eq:ccc}
    CCC = \frac{2\rho_{s,t} \sigma _s \sigma_t}{{\sigma _s}^2 {\sigma_t}^2 + {( \mu _s - \mu_t )}^2}
\end{equation}
\noindent where $\mu_s$ and $\sigma_s$ (or $\mu_t$ and $\sigma_t$) are the mean and standard deviation of the radiomic features belong to the same feature class in a synthesized (or target) image respectively, and $\rho_{s,t}$ is the Pearson correlation coefficient between $s$ and $t$. CCC ranges from -1 to 1 and is the higher the better. 

Peak Signal-to-Noise Ratio (PSNR, see Eq.~\ref{eq:error_distance}), which is a widely used metric for measuring the relative perceptual quality in image and video comparison~\cite{korhonen2012peak}. PSNR is defined as a logged ratio of the peak signal and the mean-square-error between the synthesized and the target images~\cite{korhonen2012peak} and is the higher the better.

\begin{equation}\label{eq:error_distance}
PSNR = 10. log \frac{MAX(X)^2}{\frac{1}{mn}\sum_{i=0}^{n}\sum_{j=0}^{n} (x_{ij}-x'_{ij})^2}
\end{equation}
\noindent where $X$ and $X'$ are target and synthesized images with $m\times n$ dimension respectively. All the feature values are normalized to $[0, 1]$ range.

Structural SIMilarity (SSIM, see Eq.~\ref{eq:error_ratio}) is defined as a co-relation between the synthesized and the standard images with values ranges from -1 to 1, and the value 1 indicates perfect structural similarity.

\begin{equation}\label{eq:error_ratio}
        SSIM = \frac{(2\mu_s \mu_t+c1)(2\sigma_st^2+c2)} {({\mu _s}^2 {\mu_t}^2 +c1)({\sigma _s}^2 {\sigma_t}^2 +c2)}
\end{equation}
\noindent where $\mu_s$ and $\sigma_s$ (or $\mu_t$ and $\sigma_t$) are the mean and standard deviation of the synthesized (or target) image respectively, and $\sigma_{s,t}$ is the co-variance of $s$ and $t$. $c_1$ and $c_2$ are two constant. The SSIM values ranges between -1 to 1, and the value 1 indicates prefect structural similarity.

Normalized Cross Correlation (NCC, see Eq.~\ref{eq:error_ncc}), which is the correlation between a synthesized image and its corresponding target image and is defined as:

\begin{equation}\label{eq:error_ncc}
        NCC = \frac{ \sum_{i=0}^{m}\sum_{j=0}^{n} x_{ij} x'_{ij}}{ \sqrt{\sum_{i=0}^{m}\sum_{j=0}^{n} x_{ij}^2 } \sqrt{\sum_{i=0}^{m}\sum_{j=0}^{n} {x'}_{ij}^{2} } }
\end{equation}

\noindent where $X$ and $X'$ are the target and synthesized image features respectively. All the values are normalized to $[0, 1]$.

\subsection*{Performance Evaluation} 

\begin{table*}
\footnotesize
\centering
\caption{{\bf CT image harmonization model performance comparison for images acquired with GE and Siemens scanners.} The values represent the averaged ($\pm$standard deviation) CCC score for radiomic features of the synthesized images generated using different models. }
	\begin{tabular}{| c | c | c |c | c | c | c| }
	\hline

\multicolumn{7}{|c|}{ GE to Siemens conversion }																									\\ \hline
Feature Class	&	First order			&	GLCM			&	GLDM			&	GLRLM			&	GLSZM			&	NGTDM			\\ \hline
Input	&	0.89	$\pm$	0.07	&	0.18	$\pm$	0.15	&	0.28	$\pm$	0.12	&	0.62	$\pm$	0.16	&	0.31	$\pm$	0.19	&	0.22	$\pm$	0.2	\\ \hline
CycleGAN	&	0.98	$\pm$	0.02	&	0.24	$\pm$	0.24	&	0.53	$\pm$	0.44	&	0.74	$\pm$	0.06	&	0.24	$\pm$	0.21	&	0.41	$\pm$	0.35	\\ \hline
CVH-CT$^1$	&	1.00	$\pm$	0.00	&	0.85	$\pm$	0.14	&	\textbf{0.89}	$\pm$	\textbf{0.28}	&	0.79	$\pm$	0.21	&	0.41	$\pm$	0.05	&	\textbf{0.86	$\pm$	0.18}	\\ \hline
CVH-CT$^2$	&	1.00	$\pm$	0.00	&	0.81	$\pm$	0.23	&	0.85	$\pm$	0.15	&	0.80	$\pm$	0.18	&	\textbf{0.77}	$\pm$	\textbf{0.12}	&	0.82	$\pm$	0.13	\\ \hline
CVH-CT	&	\textbf{1.00}	$\pm$	\textbf{0.00}	&	\textbf{0.88}	$\pm$	\textbf{0.11}	&	0.88	$\pm$	0.12	&	\textbf{0.90}	$\pm$	\textbf{0.14}	&	0.72	$\pm$	0.22	&	0.84	$\pm$	0.16	\\ \hline
\multicolumn{7}{|c|}{ Siemens to GE conversion}	\\ \hline
Input	&	0.89	$\pm$	0.07	&	0.018	$\pm$	0.15	&	0.28	$\pm$	0.12	&	0.62	$\pm$	0.16	&	0.31	$\pm$	0.19	&	0.22	$\pm$	0.2	\\ \hline
CycleGAN	&	0.99	$\pm$	0.01	&	0.60	$\pm$	0.15	&	0.66	$\pm$	0.18	&	0.47	$\pm$	0.12	&	0.19	$\pm$	0.18	&	0.73	$\pm$	0.26	\\ \hline
CVH-CT$^1$	&	1.00	$\pm$	0.00	&	0.82	$\pm$	0.22	&	0.81	$\pm$	0.10	&	0.83	$\pm$	0.15	&	0.79	$\pm$	0.15	&	\textbf{0.95}	$\pm$	\textbf{0.15}		\\ \hline
CVH-CT$^2$	&	1.00	$\pm$	0.00	&	0.86	$\pm$	0.12	&	0.80	$\pm$	0.11	&	0.85	$\pm$	0.10	&	\textbf{0.80}	$\pm$	\textbf{0.12}	&	0.88	$\pm$	0.11 	\\ \hline
CVH-CT	&	\textbf{1.00}	$\pm$\textbf{	0.00}	&	\textbf{0.93}	$\pm$	\textbf{0.28}	&	\textbf{0.87}	$\pm$	\textbf{0.15}	&	\textbf{0.86}	$\pm$	\textbf{0.00}	&	0.75	$\pm$	0.12	&	0.92	$\pm$	0.18	\\ \hline
	\end{tabular} \label{table:ccc}
\end{table*}

\begin{table*}[!bt]
\footnotesize
\centering
\caption{{\bf CT image harmonization model performance comparison for images acquired with GE and Siemens scanners.} The values indicate the visual image quality of the harmonized images measured using PSNR, SSIM, and NCC. }
	\begin{tabular}{| c | c | c |c | c | c |c |  }
	\hline

 &  \multicolumn{3}{|c|}{ GE to Siemens conversion }	& \multicolumn{3}{|c|}{ Siemens to GE conversion }		\\ \hline									
Metrics	&	PSNR			&	SSIM			&	NCC		&		PSNR			&	SSIM			&	NCC			\\ \hline
Input	&	24.11	$\pm$	0.62	&	\textbf{0.99	$\pm$	0.02}	&	0.84	$\pm$	0.05	&	24.11	$\pm$	0.62	&	\textbf{0.99	$\pm$	0.02}	&	0.84	$\pm$	0.05	\\ \hline
CycleGAN	&	25.58	$\pm$	0.14	&	0.99	$\pm$	0.24	&	0.93	$\pm$	0.06	&	27.60	$\pm$	0.75	&	0.60	$\pm$	0.01	&	0.90	$\pm$	0.08	\\ \hline
CVH-CT$^1$	&	26.62	$\pm$	0.28	&	0.53	$\pm$	0.44	&	0.95	$\pm$	0.05	&	27.22	$\pm$	0.50	&	0.66	$\pm$	0.02	&	0.92	$\pm$	0.11	\\ \hline
CVH-CT$^2$	&	26.50	$\pm$	0.25	&	0.62	$\pm$	0.33	&	0.95	$\pm$	0.04	&	27.15	$\pm$	0.32	&	0.70	$\pm$	0.04	&	0.92	$\pm$	0.12	\\ \hline
CVH-CT	&	\textbf{26.75	$\pm$	0.32}	&	\textbf{0.99	$\pm$	0.06}	&\textbf{	0.97	$\pm$	0.04}	&	\textbf{27.69	$\pm$	0.79}	&	\textbf{0.99	$\pm$	0.02}	&	\textbf{0.96	$\pm$	0.06}	\\ \hline

\end{tabular} \label{table:psnr}
\end{table*}

We compared CVH-CT with the conventional CycleGAN and two CVH-CT variations. CVH-CT$^1$ is constructed with CycleGAN with the proposed domain loss $\mathcal{L}_f$. CVH-CT$^2$ is constructed with CycleGAN with the proposed CBAM layers. We evaluated all the models in two settings, i.e. transform a GE image into a synthesized image with the Siemens domain, and transform a Siemens image into a synthesized image with the GE domain. Note the source image of one domain is the target image of another domain. 
Table~\ref{table:ccc} shows the effectiveness of CT image harmonization in soft-tissue ROIs. The CCC scores of synthesized and actual images are reported for six different feature classes. The row ``Input" indicates the CCC scores of the source image in the six feature classes, which serve as the baseline. The rest CCC scores indicate the performance of all the compared models regarding the improvement of feature reproducibility. In both conversions, CVH-CT achieves the best score in five feature classes. All the values in the column ``First order'' is close to 1.0 indicating that anatomic features are well preserved with all the models. The CCC scores of CVH-CT are significantly higher than that of CycleGAN on the texture feature classes.
Table~\ref{table:psnr} shows the PSNR, SSIM, and NCC scores for image quality assessment. Overall, CVH-CT effectively improved the visual quality in the synthesized images according to the three metrics. 

%
We further tested CVH-CT using unpaired image data collected from the same scanner using two reconstruction kernels, namely Bl64 and Br40. The CCC score for Br40-to-Bl64 conversion is 0.87$\pm$ 0.12 and the score for Bl64-to-Br40 conversion is 0.86$\pm$ 0.10, indicating that CVH-CT works well not only on cross-vendor image harmonization but also on cross-protocol image harmonization. 

\begin{figure}[bt!]
\centering
\includegraphics[width=.85\columnwidth]{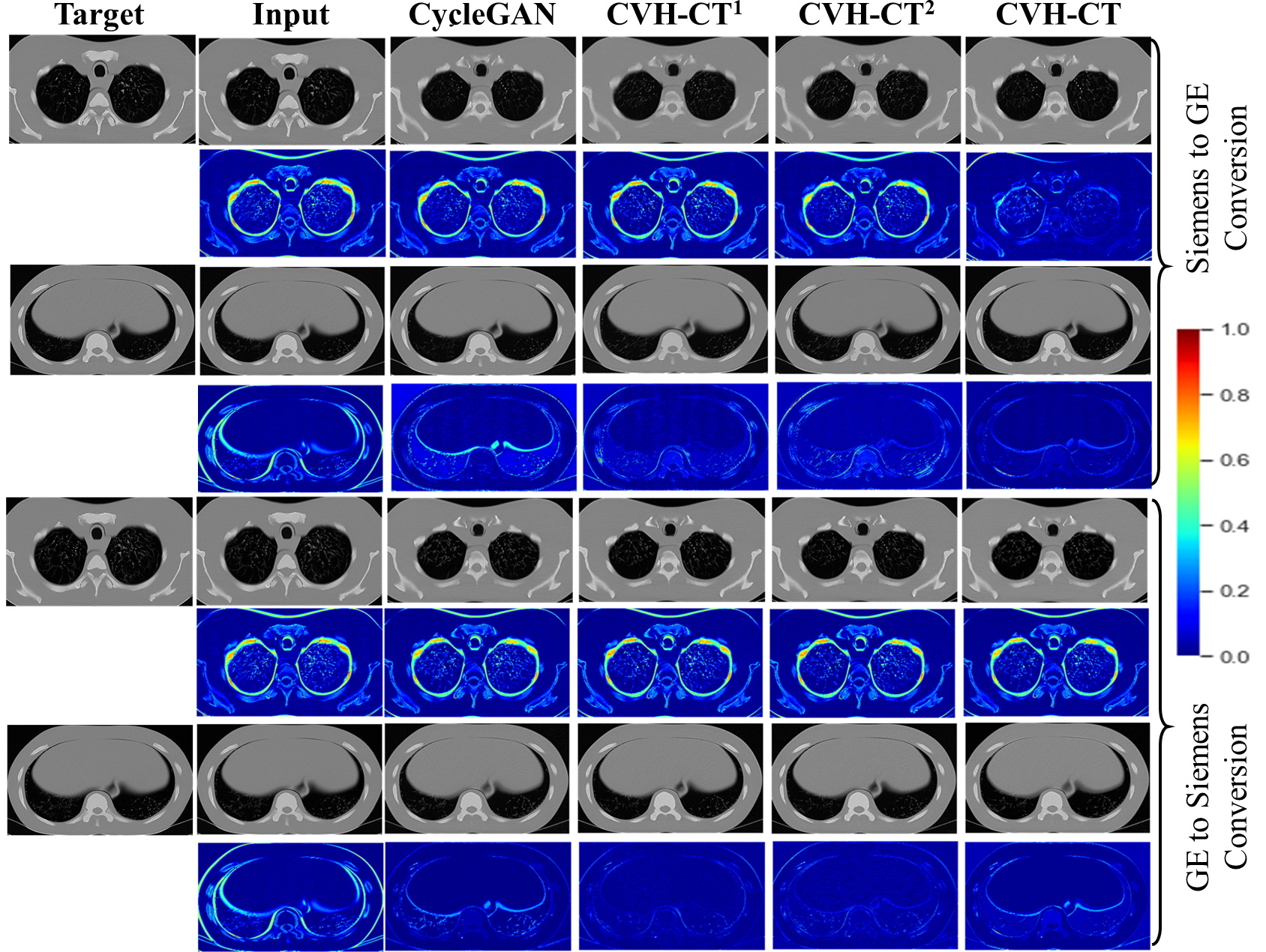}
\caption{\textbf{Case study of model performance comparison.} Heat maps of the residual images obtained by comparing two target images with the corresponding synthesized images under both GE-to-Siemens and Siemens-to-GE conversions. The input column includes images need to be harmonized and the rest of the columns are for the results of the respective models. Overall, CVH-CT has the lowest residues. Values close to 0 (blue) in heat map indicate high similarity between target and synthesized images, while values close to 1 (red) indicate high dissimilarity.}  \label{fig:residual}
\end{figure}

Figure~\ref{fig:residual} shows the results of two sample images obtained from CycleGAN, CVH-CT$^1$, CVH-CT$^2$, and CVH-CT. The source images of GE-to-Siemens conversion are the target images for Siemens-to-GE conversion and vice versa. The residual images at the second, fourth, sixth, and eighth rows were calculated as a numerical difference between a synthesized image and its corresponding target image. Values close to 0 (blue) indicate high similarity between them, while values close to 1 (red) indicate high dissimilarity between them.  Overall, CVH-CT performed clearly better than CycleGAN. The residual image of CVH-CT indicates that its synthesized images are very close to the corresponding target images. For example, in the first row, the outer side of the lung has a high mismatch in all the synthesized images except for the image from the proposed CVH-CT model.  


\begin{figure}[bt!]
\centering
\includegraphics[width=.9\columnwidth]{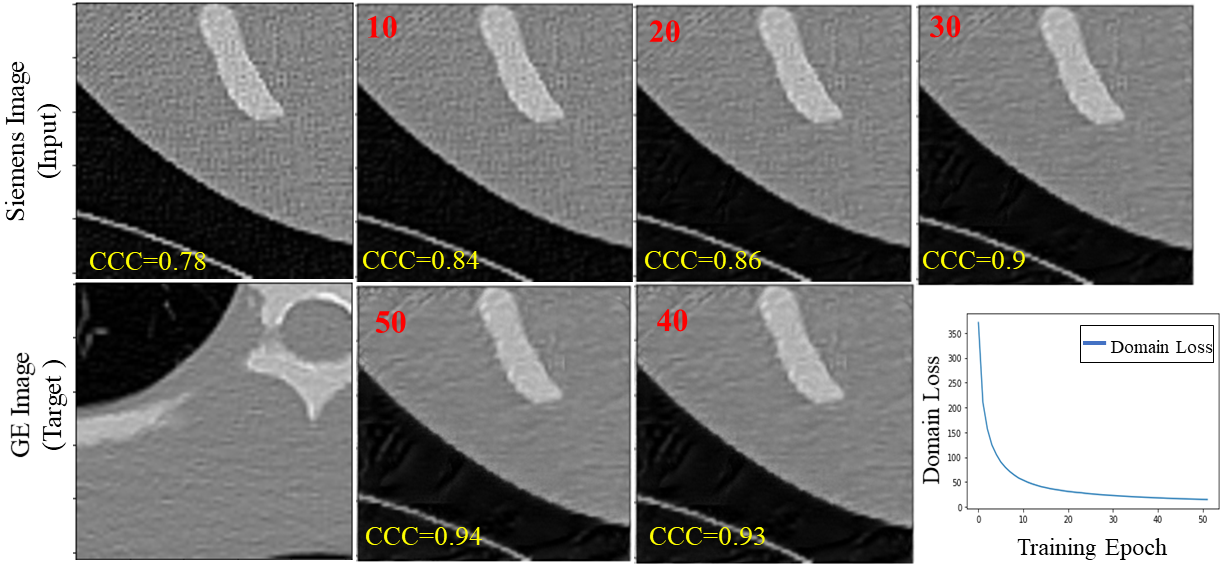}
\caption{\textbf{Case study of domain loss of unpaired CT images.} The input and target images are unpaired since they have different anatomic features, yet with the domain loss CVH-CT successfully extracts texture information from the target image. The red in the inset represents the training epoch and yellow represents the CCC score of that image. It shows that CVH-CT is able to harmonize Siemens images to fit the GE domain.} \label{fig:loss}
\end{figure}

Figure~\ref{fig:loss} shows the effectiveness of the proposed domain loss in CVH-CT.  In this case study, we synthesized an image in the GE domain using a Siemens image with different anatomic features (i.e. unpaired images). 
The visualization of images in different training epochs indicates that CVH-CT was efficiently trained to learn the textures from the GE domain. Also, these synthesized images at different epochs were not disturbed by the target image structure indicating  the proposed domain loss can learn texture details while  ignoring the structural information. The domain loss values are visualized in the right-bottom corner of the figure.

\section{Conclusions}
Data discrepancy in CT images due to the use of scanners with different imaging techniques adds an extra burden to radiologists and also creates a gap in large-scale cross-center radiomic studies. 
To facilitate large-scale medical image studies and to address the long-existing cross-center CT image data integration problem, we propose CVH-CT, a novel tool for CT image data harmonization. In CVH-CT, both the CBAM layers and the domain loss are introduced for efficient model training. CVH-CT can harmonize images without the need for paired training data.  
The experimental results demonstrate that CVH-CT is significantly better than the existing tools on CT image harmonization.  

\section*{Acknowledgements} This research is supported by NIH NCI (grant no. 1R21CA231911) and Kentucky Lung Cancer Research (grant no. KLCR-3048113817).

\bibliographystyle{unsrt}
\bibliography{camera_ready}

\end{document}